\documentclass[aps,twocolumn,longbibliography,10pt]{revtex4-1}

\usepackage{amsmath}
\usepackage[normalem]{ulem}
\usepackage{amssymb}
\usepackage{latexsym}
\usepackage{indentfirst}

\usepackage{graphicx}
\usepackage{color}
\usepackage{bm}
\usepackage{multirow}

\begin{document}

\title{Bipolar switching in an orthogonal spin transfer spin valve device}
\author{Li Ye}
\author{Georg Wolf}
\author{Daniele Pinna}
\author{Gabriel~D. Chaves}
\author{Andrew~D. Kent}
\affiliation{Department of Physics, New York University, New
             York, NY 10003, USA}
\begin{abstract}
We demonstrate current-induced bipolar switching in in-plane magnetized spin-valve devices
that incorporate a perpendicularly magnetized spin polarizing layer.
Further, hysteretic transitions into a state with intermediate resistance occur at high currents, again for both current
polarities. These transitions are shown to be consistent with a macrospin model that
considers a spin-polarized current that is tilted with respect to the free layer's plane,
due to the presence of spin-transfer torque from the polarizing layer. These unique
switching characteristics, which have their origin in the non-collinear layer magnetizations,
are of interest for magnetic random access memory and spin-torque oscillator devices.
\end{abstract}
\date{\today}

\maketitle

\section{Introduction}
Spin transfer torque (STT) devices continue to be intensively studied both for their
potential to realize current controlled devices, such as magnetic memory and
oscillators, and for fundamental interest in the nature of current induced magnetic
excitations \cite{Brataas2012}. A prototypical STT device consists of two magnetic
layers separated by a non-magnetic layer, either a metal, forming a spin-valve or a thin insulating layer,
forming a magnetic tunnel junction (MTJ). One of the magnetic layers can be excited by
spin-torques, while the other is fixed. A defining characteristic of such
devices is that for a set initial state (e.g. the layer magnetizations aligned either
parallel or antiparallel) current induced switching only occurs for one current polarity \cite{Katine2000}.
Another characteristic of such devices is that the spin-transfer torques are small
when the layers are nearly collinearly magnetized leading to stochastic switching
(see, for example, \cite{Liu2014}).

Here we consider STT devices that incorporate an orthogonally magnetized spin-polarizing
layer in an in-plane magnetized spin-valve, known as orthogonal spin-transfer (OST) device
\cite{Kent2004}. A polarizing layer that is magnetized perpendicular to the free layer
can significantly improve write speed and energy efficiency of STT-magnetic random access
memories \cite{Liu2010,Nikonov2010,Lee2011,Rowlands2011,Liu2012,Park2013,Ye2014}
by providing a large initial spin-transfer torque (i.e. a large torque the moment a current
is applied). An OST device can also function as a microwave oscillator, because the
polarizer can produce precessional magnetization dynamics, with the free layer precessing
out of the film plane about an axis normal to the layer planes~\cite{Houssameddine2007,Ebels2008,
Firastrau2008}. This precessional motion also can be used for ultrafast magnetization switching. For instance,
sub-nanosecond switching has been observed in OST-MTJ devices that incorporate a
perpendicularly magnetized polarizing layer~\cite{Liu2010,Rowlands2011,Liu2012}.
Further, for nanosecond pulses the switching has been observed to be bipolar and
to induce magnetization precession \cite{Liu2010,Liu2012}. However, deterministic bipolar
switching between P and AP states has not been observed in quasistatic measurements.
Further, the effect of applied fields on these current-induced switching thresholds has not been
reported or considered in model studies.
\begin{figure}[t]
  \begin{center}
   \includegraphics[width=3in,keepaspectratio=True]
   {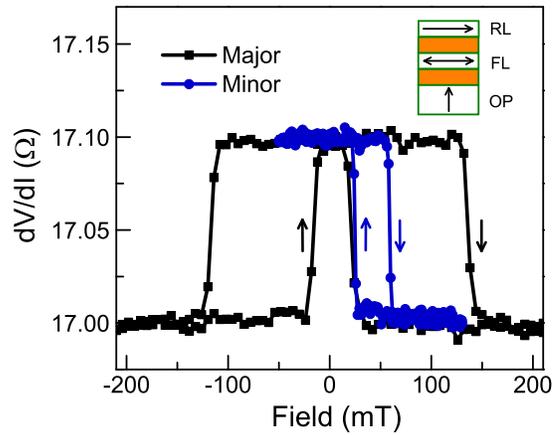}
  \end{center}
\caption{Resistance versus in-plane applied field hysteresis loops. The major loop (black curve)
shows the switching of both the free and reference layers. The minor loop (blue curve) shows the
response of just the free layer. The loop is centered at $41$ mT due to dipolar interactions between
the reference and free layer. Inset: Schematic of the spin-valve's layer stack showing the out-of-plane
magnetized polarizing layer (OP), in-plane magnetized  free layer (FL) and reference layer (RL).}
\label{Fig1}
\end{figure}

Here we report the observation of bipolar switching of OST spin valve devices,
where in contrast to conventional collinearly magnetized SV devices,
the switching from parallel (P) to antiparallel (AP) resistance states and the reverse transition (AP to P)
occurs for both current polarities in a range of applied fields.  Further, we find hysteretic transitions into an
intermediate resistance state (IR) at large current, with the IR state persisting to currents less than the threshold
currents for P to AP and AP to P switching.  A macrospin model, including spin transfer torques
from the reference and polarizing layers as well as finite
temperature effects, captures the hysteretic current-driven transitions observed in experiment.

\section{Experiment}
The layer stacks consist of a perpendicularly magnetized spin-polarizing layer, a
non-magnetic metallic spacer layer, a free magnetic layer followed by another
non-magnetic metallic spacer layer and a reference magnetic layer, as illustrated
in the inset of Fig.~\ref{Fig1}. The polarizer consists of a Co/Pd and Co/Ni multilayer,
with the Co/Ni multilayer closest to the free layer (FL), providing a highly spin-polarized
current \cite{Liu2010}. The FL is a 3 nm thick CoFeB layer. The full layer stack is
6.2 [Co/Pd][Co/Ni]/10 Cu/3 CoFeB/10 Cu/12 CoFeB, with the layer thicknesses indicated
in nanometers. The stack was patterned into nanopillar devices with various shapes and sizes
using e-beam lithography and ion-milling. Here we present results on 50~nm $\times$ 100~nm
devices in the shape of an ellipse. The magnetic easy axis of free layer is in the film
plane along the long axis of ellipse due to magnetic shape anisotropy. Shape anisotropy
also sets the magnetization direction of the $12$~nm thick CoFeB reference layer (RL).

Figure 1 shows measurements of the differential resistance ($dV/dI$)
as a function of applied field along the easy axis. The measurements
are made with a lock-in amplifier using an ac current of
$200$~$\mu$A at a frequency of $473$~Hz. A field sweep from $-200$
mT to $200$ mT (major hysteresis loop) shows steps in resistance of
$0.1~\Omega$ indicative of switching of the FL from P to AP relative
to the RL. The coercive field of the RL is about $150$ mT. A minor
loop ($-50$~mT to $140$~mT) shows the switching of only the FL. The
change in resistance between P and AP states is $\Delta R_{AP-P}=0.1
~\Omega$. The coercive field for AP to P FL transitions is
$H_c^+=59$~mT and the coercive field for P to AP FL transitions is
$H_c^-=23$~mT. The minor loop is centered at $H_0=(H_c^++H_c^-)/2=41$~mT
due to dipolar coupling between the FL and RL. Thus an external field of
$H_0$ corresponds (on average) to zero effective field applied to the FL.
\begin{figure}[t]
  \begin{center}
   \includegraphics[width=3in,
    keepaspectratio=True]
   {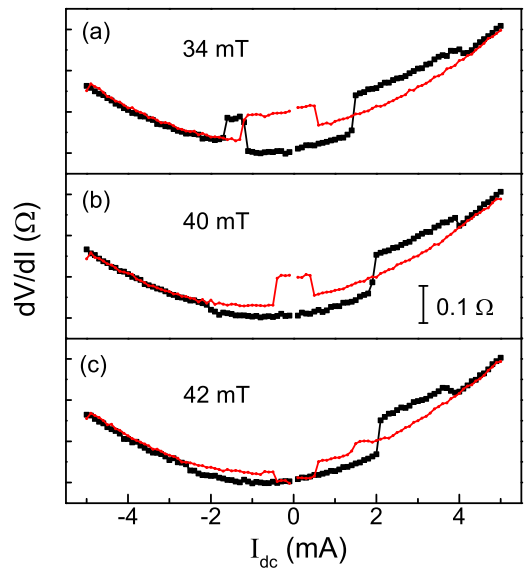}
  \end{center}
  \caption{Differential resistance versus current at various easy axis applied fields starting from
the P state. The magnitude of the current $|I_{dc}|$ is increased
(black curves) and then decreased (red curves). (a) The black curve
shows switching from P to AP at 1.5 mA and also -1.2 mA, i.e. the
switching occurs for both polarities of the current. At larger
positive and negative current the resistance change is intermediate
of that of the P to AP transitions. On reducing the current there is
a transition from the intermediate resistance (IR) state into an AP
state. (b)  At $40$ mT switching from P to AP only occurs for
positive polarity current and on reducing the current there is an IR
to AP state transition for $|I_{dc}|\lesssim 1$ mA. (c) At $42$ mT
switching from P to AP again only occurs for positive polarity
current. However, on reducing the applied field the transition is
from IR to P for $|I_{dc}|\lesssim 1$ mA.} \label{Fig2}
\end{figure}

Current induced switching was characterized by measuring the
differential resistance as a function of current for a
series of easy axis applied fields. The magnetic state (P or AP) is
first set by applying a large magnetic field $200$~mT and then a
lower field ($\geq25$~mT for the P state, $\leq58$~mT for the AP state).
Then the current $I_{dc}$ was slowly ramped ($\simeq 0.1$~mA/s) from $0$ to $\pm 5$~mA
and then ramped back to $0$~mA, with $dV/dI$ versus $I$ recorded at each measuring field.
Positive current corresponds to electron flow from polarizing to the reference layer (from
bottom to the top of the layer stack represented in the inset of Fig.~\ref{Fig1}).
In this case the spin-torque associated with the RL spin-torque favors an AP state
for positive current. Representative measurement results
starting from the P state are shown in Fig.~\ref{Fig2}. Similar
results were found in measurements starting from the AP state, which
are discussed below.

In Fig.~\ref{Fig2}(a) as the magnitude of the current is increased (black curves), there is a
discrete increase in differential resistance of $\Delta R_{AP-P}=0.1~\Omega$, associated
with a P to AP transition. This is seen to occur for both polarities of the current. On further
increasing the current there is a change in resistance of about half of $\Delta R_{AP-P}$, i.e. a transition
into an intermediate resistance state (IR). On decreasing the current (red curves) the resistance
eventually returns to that of the device's AP state. Near zero effective applied field ($H \simeq H_0$),
P to AP switching is only seen at positive current polarity (Fig. 2 (b) and (c)). Whereas, for negative current, only P to
IR transitions occur as the current is increased. When the magnitude of the current is decreased,
there is a transition from an IR state to an AP state for applied magnetic fields smaller than $H_0\;(=41$ mT)
and to a P state for fields larger than $H_0$.

\begin{figure}[t]
\begin{center}
\includegraphics[width=2.8in,keepaspectratio=True]{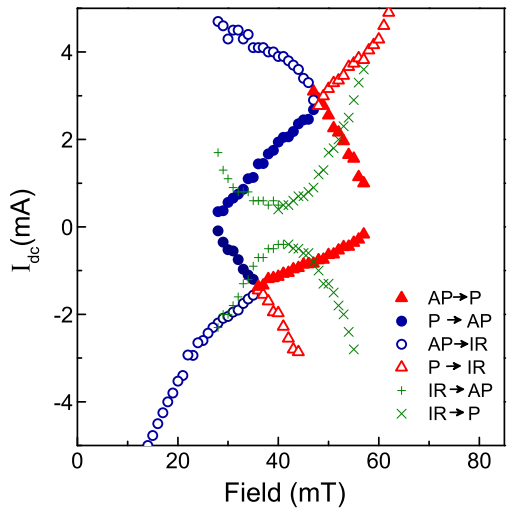}
\end{center}
\caption{Current swept state diagram of an OST spin valve device
showing the threshold currents for switching as a function of applied
easy axis field. $I_{c}^{P-AP}$ and $I_{c}^{AP-IR}$ are labeled by
solid and open blue symbols. $I_{c}^{AP-P}$ and $I_{c}^{P-IR}$ are
labeled by solid and open red symbols. The green curves indicate the
$I_{c}^{IR-AP}$ (crosses) and $I_{c}^{IR-P}$ (dashes), showing the
bistability range of the IR states.} \label{Fig3}
\label{ExpStateDiagram}
\end{figure}

This seemingly complex switching behavior can be summarized by
plotting the threshold currents for switching between resistance
states in a current-applied field state diagram (Fig.~\ref{Fig3}).
Each symbol in this diagram corresponds in a discrete change in the
resistance. The solid symbols represent resistance changes of
$\Delta R_{AP-P}$ corresponding to transitions between P and AP
states. They form a diamond-shaped central zone within which both P
and AP states are possible. When the current is greater than
$2.9$~mA or is less than $-1.4$~mA, the step change in resistance is
less than $\Delta R_{AP-P}$. The boundaries are labeled by open
symbols and correspond to P and AP to IR transitions. These
boundaries meet and join the P to AP transition boundaries. Further,
they define two triangular zones that encompass IR states at high
current magnitudes (both for positive and negative current
polarities). As the current is swept back to zero, two parabolic
shaped curves (green) show the IR to P or AP transitions.

The general features of the state diagram of an OST-SV device
are the following: (1) For magnetic fields near the FL coercive fields ($H_c^+$ and $H_c^-$),
current induced switching is bipolar. For fields close to but less
than $H_c^+$, AP to P transitions occur for both positive and
negative currents and for fields near but greater than $H_c^-$, P
to AP transitions occur for both current polarities. (2) Near $H_0$,
the center of the FL's hysteresis loop, the switching occurs for
only one current polarity, positive current for the P to AP
transition and negative current for the AP to P current. (3) At
large currents, transitions into an IR state are observed and this
state persists even as the current is reduced well below the
threshold current for P/AP transitions. These features were
seen in all ten $50$~nm $\times$ $100$~nm ellipse devices
that we studied.

\begin{figure}[t]
  \begin{center}
   \includegraphics[width=3.5in,
    keepaspectratio=True]
   {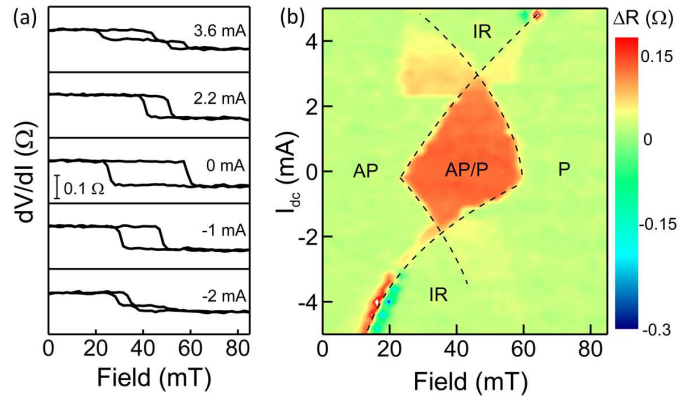}
  \end{center}
  \caption{(a) Representative FL minor hysteresis loops measured
with a slowly swept field at several fixed currents. The  scale bar
shows $\Delta R_{AP-P}=0.1~\Omega$, the resistance difference
between the AP and P states. (b) State diagram constructed from
$dV/dI-H$ hysteresis loops. The color represents $\Delta R$, the
resistance difference between field increasing and field decreasing
measurements. The central zone (orange color) corresponds to the
AP/P bistable zone. Black dashed curves trace the boundaries between
P, AP and IR states.} \label{Fig4}
\end{figure}

The device states can also be determined by measuring the
differential resistance as a function of field at constant current.
This is shown in Fig.~\ref{Fig4}. Fig.~\ref{Fig4}(a) shows
representative hysteresis loops at several currents. At zero dc
current the coercive field ($H_c=(H_c^+-H_c^-)/2$) of the FL is
largest and the coercive field decreases as the current magnitude
increases. For currents greater than $2.9$ mA or less than $-1.4$
mA, a plateau at a resistance between that of the P and AP state
resistances is seen, with the field range of the plateau increasing
with current magnitude.

A field swept state diagram is constructed as follows. The
resistance measured with decreasing field is subtracted from the
resistance measured with increasing field. The resistance difference
$\Delta R$ is then plotted on a color scale versus current and field
(Fig.~\ref{Fig4}(b)). $\Delta R$ is nonzero only in field ranges in
which the device response is hysteretic. The boundaries between zero
and and non-zero $\Delta R$ regimes are the boundaries between the
P, AP and IR states. Thus the same general switching characteristics
are observed in both current and field swept measurements.

\section{Model}
To understand the device switching characteristics, we
consider the spin-transfer torques acting on the FL along with its magnetic anisotropy in a macrospin
model. The dynamics of $\mathbf{m}$, a unit vector in the magnetization direction of the free layer ($\mathbf{m}=\mathbf{M}/M_s$),
is  given by the Landau-Lifshitz-Gilbert-Slonczewki (LLGS) equation:
\begin{equation}
\frac{d\mathbf{m}}{d\tau}=-\mathbf{\Gamma}_{\mathrm{LLG}}+\mathbf{\Gamma}_{\mathrm{th}}+\mathbf{\Gamma}_{\mathrm{S}},
\label{LLGS}
\end{equation}
where $\mathbf{\Gamma}_{\mathrm{LLG},\mathrm{th},\mathrm{S},}$
represents the LLG, thermal torque and spin-torque. The LLG torque is given by
$\mathbf{\Gamma}_{\mathrm{LLG}}=-\mathbf{m}\times\mathbf{h}_{\mathrm{eff}}-\alpha\mathbf{m}\times(\mathbf{m}\times\mathbf{h}_{\mathrm{eff}})$,
with effective field $\mathbf{h}_{\mathrm{eff}}=\frac{-1}{\mu_0 M_s^2V}\nabla_{\mathbf{m}}U(\mathbf{m})$, volume of the magnetic element $V$
and damping constant $\alpha$. Time in Eqn.~\ref{LLGS}
has been normalized by the precession frequency, $\gamma \mu_0 M_s$ (i.e. $\tau=\gamma \mu_0 M_st$), where $\gamma$ is the gyromagnetic
ratio. The thermal torque  $\mathbf{\Gamma}_{\mathrm{th}}$ is induced by a Gaussian distributed random field $\mathbf{h}_{\mathrm{th}}$~\cite{Pinna2013}. The FL
has a biaxial magnetic anisotropy energy:
\begin{equation}
U(\mathbf{m})=U_0\left(Dm_z^2-m_x^2\right),
\end{equation}
with an easy axis along $\hat{x}$ and hard axis along $\hat{z}$, where $D$ is the ratio of the hard to easy axis anisotropy $D=2M_s/H_K$ and
$U_0=\frac{1}{2}\mu_0M_sH_KV$ is the energy barrier to  magnetization reversal.

Spin-torque contributions due to both the polarizer (magnetized out-of-the film plane, along $\hat{z}$) and RL (magnetized in the film plane, along $\hat{x}$) can be described in terms of effective spin-polarization direction that is tilted with respect to the plane:
\begin{equation}
\mathbf{\Gamma}_{\mathrm{S}}=\tilde{I}\mathbf{m}\times(\mathbf{m}\times\mathbf{n}_S)\nonumber,
\end{equation}
\begin{equation}
\mathbf{n}_S=\frac{\eta_R}{1-\lambda_Rm_x}\mathbf{\hat{x}}+\frac{\eta_P}{1-\lambda_Pm_z}\mathbf{\hat{z}}.
\label{STT}
\end{equation}
Here $\eta_{R,P}$ and $\lambda_{R,P}$ are the spin polarizations and spin-torque asymmetry parameters for the RL and OP, respectively. $\tilde{I}=(\hbar/2e)I/(\mu_0M_s^2V)$ is a normalized applied current. Therefore, the
combined spin-torque acting on the FL magnetization can be written as:
\begin{equation}
\mathbf{\Gamma}_{\mathrm{S}}=\tilde{I}\frac{\sqrt{1+\tan^2(\omega_{\mathrm{eff}})}}{1-\lambda_Rm_x}\mathbf{m}\times(\mathbf{m}\times\mathbf{\hat{n}}),
\end{equation} where the orientation of the effective spin-polarization axis
$\mathbf{\hat{n}}$ is tilted with respect to the in-plane (IP) direction by an
angle $\omega_{\mathrm{eff}}$ with
\begin{equation}
\tan(\omega_{\mathrm{eff}})=\frac{\eta_P}{\eta_R}\cdot\frac{1-\lambda_Rm_x}{1-\lambda_Pm_z}=\tan(\omega)\frac{1-\lambda_Rm_x}{1-\lambda_Pm_z}.
\end{equation} Naturally, in the case of $\eta_P=0$ (no out-of-plane polarizer), $\mathbf{\Gamma}_{\mathrm{S}}$
will reduce to the conventional collinear spin-torque expression and
$\mathbf{\hat{n}}$ will lie in plane.

A qualitative understanding of central zone of the state-diagram can be seen from the form of the spin-transfer torque in Eqs.~\ref{STT}.
The torque associated with the reference layer is initially collinear with the damping torque. It thus leads to switching via the
antidamping mechanism, typical of spin-transfer devices with collinear magnetizations. However, the spin-transfer torque from the polarizer ($\propto
\mathbf{m}\times(\mathbf{m}\times\hat{z})$) is equivalent to an effective field in the direction $\mathbf{m}\times\hat{z}$, which is initially in the direction of the FL's medium axis $\hat{y}$.
Such a field reduces the FL's easy axis coercive field (for both current polarities), as is the case in the Stoner-Wohlfarth model with a medium axis magnetic field. In the Stoner-Wohlfarth model the result is an astroid shaped switching boundary, which resembles the diamond shaped bistable central zone of our state diagram.

More quantitatively, the spin-torque asymmetry parameters $\lambda_{R,P}$ lead to torques
that depend on the magnetization state of the FL. For example, different current magnitudes are typically necessary
for AP to P and P to AP switching \cite{Slonczewski1996}. Generally, larger currents are needed for P to AP switching at the same field, as seen in the state diagram (Fig.~\ref{ExpStateDiagram}). For simplicity in our model, we consider $\lambda_P=0$, as $m_z$ is typically small during the switching process. Therefore, $\lambda_R$ accounts for the main asymmetries we observe in spin-torque switching.
\begin{figure}[t]
  \begin{center}
   \includegraphics[width=2.8in,
    keepaspectratio=True]
   {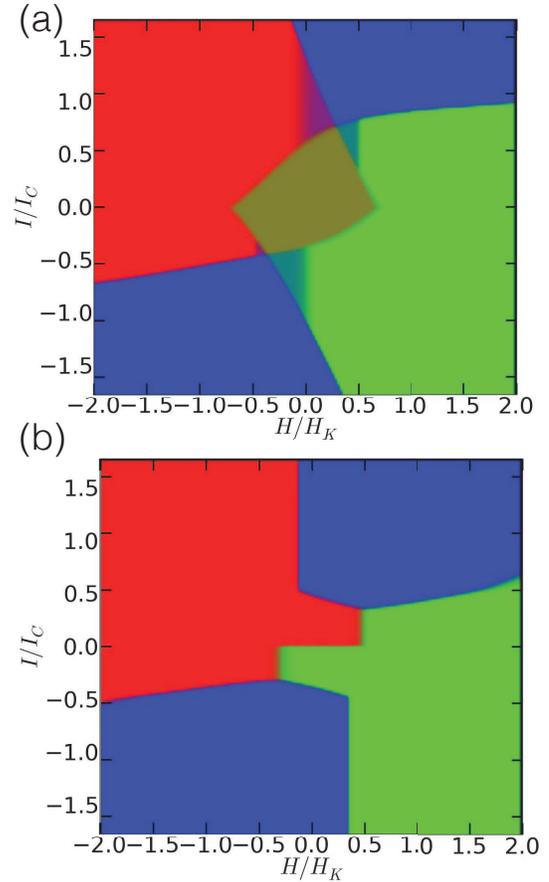}
  \end{center}
  \caption{Simulations of an ensemble of 5000 macrospins represented as a state-diagram
with current increasing (a) and current decreasing (b). The three relevant
states AP, P and IR, are color coded as red, green and blue
respectively. Each data point is represented by a RGB color that is
determined by the proportion of the ensemble populating each
corresponding state. Currents are shown in units of switching positive
current at zero field and room temperature. Applied fields are shown in
units of anisotropy field. The parameters used in the simulation are described
in the main text.}
\label{Fig5}
\end{figure}
We study the switching by simulating an ensemble
of 5000 macrospins under the influence of spin-torque, thermal noise
and an applied field. The ensemble is initially taken to be
thermalized in one of its (two possible) equilibrium easy-axis states before a current
is applied. After a current is applied the ensemble reaches a new steady-state
configuration over the course of a microsecond. The simulation is then
repeated at a higher current in incremental steps using the steady-state ensemble of the
previous current as the initial condition. Upon reaching
the limit of our current range, we incrementally reduce the current to reproduce the procedure in the
current swept experiments.

Simulation results are plotted in Fig. 5 both for current ramped-up
(a), and current ramped-down (b) cases, with parameters determined as follows.
$D$ is governed by magnetic shape anisotropy and is calculated based on the FL's shape to be $17$.
The spin-torque asymmetry is taken to reproduce the measured ratio of positive $I_c^+$ to negative $I_c^-$ switching
currents at effective zero field $I_c^+/I_c^-=2$, giving $\lambda_{R}=0.5$.
$U_0=\frac{1}{2}\mu_0M_sH_KV$ is estimated to be $3.3 \times 10^{-19}$~J (i.e. $U_0/k_BT=80$, with $T=300$~K),
taking $\mu_0 M_s=1.5~T$ and $\mu_0H_K=35$~mT.  We then ran simulations with spin torque ratios ${\eta_P}/{\eta_R}=0, 0.24, 0.51, 0.68$
and $\alpha=0.04$.
The results are shown as a colormap where red and green represent in-plane AP/P configurations and blue
corresponds to IR states. Fig.~\ref{Fig5} shows results for ${\eta_P}/{\eta_R} = 0.68$.
(Results for the other spin-torque ratios studies are shown in the Supplementary Materials.)
Depending on the proportions in which the
ensemble is partitioned between the three available states,
bistability regions arise and are represented by a superposition of
the colors, an example being the dark yellow (green+red) P/AP
bistability region at the center of the state diagram. The magnetic
field is normalized to $H_K$ and current is
normalized by the positive critical current $I_{c}^+$ at zero
temperature. The normalized critical current
($I_{c}^{+}/I_c$, $I_{c}^{-}/I_c$) and coercive field ($H_{c}^{+}/H_K$,
$H_{c}^{-}/H_K$) are smaller than 1 because of thermal
fluctuations.

The simulation captures the main switching features observed in the experiment
both for current ramping up and down. First, we observe a distorted diamond shaped
central P/AP bistable central zone (Fig.~\ref{Fig5}(a)), which shows
bipolar switching near the layer's coercive field. The distortion (e.g. the lower
switching current for AP to P transitions) is associated with the non-zero spin-torque asymmetry parameter
$\lambda_R$.  Second, there are transitions into an out-of-plane precessional state
which we associate with the IR state we observe experimentally.  We find that the threshold current
for P/AP to IR transitions is higher when the current is increasing than the IR to P/AP
transitions when the current is decreasing (Fig.~\ref{Fig5}(b)), as observed
in experiment. The vertical boundaries in Fig.~\ref{Fig5}(b) are
artifacts that result from not having applied large enough
currents to realize an IR state when the current was increasing. In this case,
when decreasing the current, the ensemble appears to not
transition out of its initial AP or P configuration. Deviations between the
experiment and macrospin model appear at large currents. For example, the curvature
of the AP to IR switching boundary is positive in the simulation but negative
in the experimental data. This indicates that more sophisticated models, such as micromagnetic models, of the
magnetization dynamics may be needed to explain the large current dynamics.

The hysteretic transitions to the IR state can be understood in a recent analytic theory
which examined the influence of the spin-torque ratio (${\eta_P}/{\eta_R}$) on the magnetization
dynamics ~\cite{Pinna2013, Pinna2014}. The theory considered a zero field case in the absence
of spin-torque asymmetries (i.e. $\lambda_{R,P}=0$, $\omega_{\mathrm{eff}}=\omega$).
A key result was that for spin-polarization tilts ($\omega$) larger than a critical value there are
hysteric transitions into a stable out-of-plane precessional (OOP) state.
The condition is $\omega>\omega_C=\mathrm{tan}^{-1}(1/\sqrt{D})$. In this
case a critical threshold current $\tilde{I}^{OOP}=(2/\pi)\sqrt{1+D}/\sin{\omega}$ exists above
which an OOP orbit is stable. However, a larger current $\tilde{I}^{IP}=\sqrt{D}\tilde{I}^{OOP}$ must
be applied to establish an OOP state starting from a P or AP state. This analytic theory thus
predicts the hysteretic transition between IR and P/AP states that are observed in this
experiment and also shows the material parameters that determine
the magnitude of the hysteresis.

\section{Summary}
In summary, we have systematically studied current-induced transitions in OST-SV devices and determined a
field-current state-diagram for these transitions, which involve switching between three
different resistance states, P, AP and IR. Deterministic bipolar switching between
P and AP states is observed in a range of fields near the FL coercive field, resulting in
a distinct diamond shaped AP/P bistable state regime in field-current state-diagram.
The main features are understood in a macrospin model that considers the effective
spin-torque acting on the free layer as a superposition of the torques from the OP
and RL. This simple model thus provides guidance in understanding and optimizing the switching
characteristics of OST-devices. These results also show the unique spin-transfer switching
characteristics of devices with non-collinear magnetizations.

\section*{Acknowledgements}
This research was supported by NSF-DMR-1309202 and in part by IARPA and SPAWAR contract N66001-12-C-2019.

\newpage
\begin{widetext}
\begin{figure*}[!t]
\begin{center}
   \includegraphics[width=7.0 in,keepaspectratio=True]
   {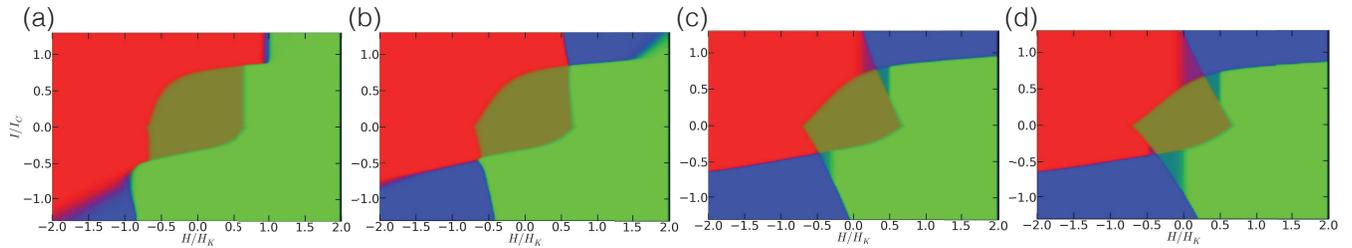}
  \end{center}
  \caption{Simulations of state diagrams for different ratios of the spin torque from the out-of-plane polarizer and in-plane reference layer: (a) ${\eta_P}/{\eta_R}$ = 0, (b) 0.24, (c) 0.51, (d) and 0.68. (d) Closely reproduces the experimental data, as described in the main text. Red: AP state, green: P state, dark yellow: AP/P bistable state.}
\label{Fig1SM}
\end{figure*}
\end{widetext}

\section{Supplementary Material}

As discussed in the main text, simulations of an ensemble of macrospins were conducted for a range spin torque ratios, ${\eta_P}/{\eta_R}$. Here we show the effect of varying the spin-torque ratios on the state diagrams and device switching characteristics.

The simulations were done with parameters: $D=17$, $\lambda_{R}=0.5$, $\lambda_{P}=0$, $\mu_0H_K=35$~mT, $M_S=1.5$~T, (i.e. $U_0/kT=80$, $T=300$~K) and damping $\alpha=0.04$. Results are shown in Fig.~\ref{Fig1SM} for (a) ${\eta_P}/{\eta_R}=0$, (b) $0.24$, (c) $0.51$ and (d) 0.68. As described in the main text, the states AP, P and IR, are indicated in red, green and blue colors
respectively. ${\eta_P}/{\eta_R}=0$ corresponds to no out-of-plane polarizer. The diagram in Fig.~\ref{Fig1SM}(a) thus shows the switching boundaries for a purely collinearly magnetized device.  There are two vertical switching boundaries and the switching is unipolar: P to AP switching occurs only for positive current while AP to P switching occurs only for negative currents. With increasing spin-torque associated with the out-of-plane magnetized polarizer the initially vertical switching boundaries acquire a negative slope. In case (c) the switching for AP to P and P to AP can now occur for both current polarities in a range of applied fields. The state diagram in (d) with ${\eta_P}/{\eta_R}=0.68$ shows the diamond shape we observe in experiment. The results also show that including an out-of-plane polarizer torque leads to a reduced area of the P/AP bistable central zone and also that the switching currents are reduced.


\begin{thebibliography}{17}%
\makeatletter
\providecommand \@ifxundefined [1]{%
 \@ifx{#1\undefined}
}%
\providecommand \@ifnum [1]{%
 \ifnum #1\expandafter \@firstoftwo
 \else \expandafter \@secondoftwo
 \fi
}%
\providecommand \@ifx [1]{%
 \ifx #1\expandafter \@firstoftwo
 \else \expandafter \@secondoftwo
 \fi
}%
\providecommand \natexlab [1]{#1}%
\providecommand \enquote  [1]{``#1''}%
\providecommand \bibnamefont  [1]{#1}%
\providecommand \bibfnamefont [1]{#1}%
\providecommand \citenamefont [1]{#1}%
\providecommand \href@noop [0]{\@secondoftwo}%
\providecommand \href [0]{\begingroup \@sanitize@url \@href}%
\providecommand \@href[1]{\@@startlink{#1}\@@href}%
\providecommand \@@href[1]{\endgroup#1\@@endlink}%
\providecommand \@sanitize@url [0]{\catcode `\\12\catcode `\$12\catcode
  `\&12\catcode `\#12\catcode `\^12\catcode `\_12\catcode `\%12\relax}%
\providecommand \@@startlink[1]{}%
\providecommand \@@endlink[0]{}%
\providecommand \url  [0]{\begingroup\@sanitize@url \@url }%
\providecommand \@url [1]{\endgroup\@href {#1}{\urlprefix }}%
\providecommand \urlprefix  [0]{URL }%
\providecommand \Eprint [0]{\href }%
\providecommand \doibase [0]{http://dx.doi.org/}%
\providecommand \selectlanguage [0]{\@gobble}%
\providecommand \bibinfo  [0]{\@secondoftwo}%
\providecommand \bibfield  [0]{\@secondoftwo}%
\providecommand \translation [1]{[#1]}%
\providecommand \BibitemOpen [0]{}%
\providecommand \bibitemStop [0]{}%
\providecommand \bibitemNoStop [0]{.\EOS\space}%
\providecommand \EOS [0]{\spacefactor3000\relax}%
\providecommand \BibitemShut  [1]{\csname bibitem#1\endcsname}%
\let\auto@bib@innerbib\@empty
\bibitem [{\citenamefont {Brataas}\ \emph {et~al.}(2012)\citenamefont
  {Brataas}, \citenamefont {Kent},\ and\ \citenamefont {Ohno}}]{Brataas2012}%
  \BibitemOpen
  \bibfield  {author} {\bibinfo {author} {\bibfnamefont {A.}~\bibnamefont
  {Brataas}}, \bibinfo {author} {\bibfnamefont {A.~D.}\ \bibnamefont {Kent}}, \
  and\ \bibinfo {author} {\bibfnamefont {H.}~\bibnamefont {Ohno}},\ }\bibfield
  {title} {\enquote {\bibinfo {title} {Current-induced torques in magnetic
  materials},}\ }\href {\doibase 10.1038/NMAT3311} {\bibfield  {journal}
  {\bibinfo  {journal} {Nature Materials}\ }\textbf {\bibinfo {volume} {11}},\
  \bibinfo {pages} {372} (\bibinfo {year} {2012})}\BibitemShut {NoStop}%
\bibitem [{\citenamefont {Katine}\ \emph {et~al.}(2000)\citenamefont {Katine},
  \citenamefont {Albert}, \citenamefont {Buhrman}, \citenamefont {Myers},\ and\
  \citenamefont {Ralph}}]{Katine2000}%
  \BibitemOpen
  \bibfield  {author} {\bibinfo {author} {\bibfnamefont {J.~A.}\ \bibnamefont
  {Katine}}, \bibinfo {author} {\bibfnamefont {F.~J.}\ \bibnamefont {Albert}},
  \bibinfo {author} {\bibfnamefont {R.~A.}\ \bibnamefont {Buhrman}}, \bibinfo
  {author} {\bibfnamefont {E.~B.}\ \bibnamefont {Myers}}, \ and\ \bibinfo
  {author} {\bibfnamefont {D.~C.}\ \bibnamefont {Ralph}},\ }\bibfield  {title}
  {\enquote {\bibinfo {title} {Current-driven magnetization reversal and
  spin-wave excitations in {C}o/{C}u/{C}o pillars},}\ }\href@noop {} {\bibfield
   {journal} {\bibinfo  {journal} {Physical Review Letters}\ }\textbf {\bibinfo
  {volume} {84}},\ \bibinfo {pages} {3149--3152} (\bibinfo {year}
  {2000})}\BibitemShut {NoStop}%
\bibitem [{\citenamefont {Liu}\ \emph {et~al.}(2014)\citenamefont {Liu},
  \citenamefont {Bedau}, \citenamefont {Sun}, \citenamefont {Mangin},
  \citenamefont {Fullerton}, \citenamefont {Katine},\ and\ \citenamefont
  {Kent}}]{Liu2014}%
  \BibitemOpen
  \bibfield  {author} {\bibinfo {author} {\bibfnamefont {H.}~\bibnamefont
  {Liu}}, \bibinfo {author} {\bibfnamefont {D.}~\bibnamefont {Bedau}}, \bibinfo
  {author} {\bibfnamefont {J.Z.}\ \bibnamefont {Sun}}, \bibinfo {author}
  {\bibfnamefont {S.}~\bibnamefont {Mangin}}, \bibinfo {author} {\bibfnamefont
  {E.E.}\ \bibnamefont {Fullerton}}, \bibinfo {author} {\bibfnamefont {J.A.}\
  \bibnamefont {Katine}}, \ and\ \bibinfo {author} {\bibfnamefont {A.D.}\
  \bibnamefont {Kent}},\ }\bibfield  {title} {\enquote {\bibinfo {title}
  {Dynamics of spin torque switching in all-perpendicular spin valve
  nanopillars},}\ }\href@noop {} {\bibfield  {journal} {\bibinfo  {journal}
  {Journal of Magnetism and Magnetic Materials}\ }\textbf {\bibinfo {volume}
  {358-359}},\ \bibinfo {pages} {233 -- 258} (\bibinfo {year}
  {2014})}\BibitemShut {NoStop}%
\bibitem [{\citenamefont {Kent}\ \emph {et~al.}(2004)\citenamefont {Kent},
  \citenamefont {Ozyilmaz},\ and\ \citenamefont {del Barco}}]{Kent2004}%
  \BibitemOpen
  \bibfield  {author} {\bibinfo {author} {\bibfnamefont {A.~D.}\ \bibnamefont
  {Kent}}, \bibinfo {author} {\bibfnamefont {B.}~\bibnamefont {Ozyilmaz}}, \
  and\ \bibinfo {author} {\bibfnamefont {E.}~\bibnamefont {del Barco}},\
  }\bibfield  {title} {\enquote {\bibinfo {title} {Spin-transfer-induced
  precessional magnetization reversal},}\ }\href@noop {} {\bibfield  {journal}
  {\bibinfo  {journal} {Applied Physics Letters}\ }\textbf {\bibinfo {volume}
  {84}},\ \bibinfo {pages} {3897--3899} (\bibinfo {year} {2004})}\BibitemShut
  {NoStop}%
\bibitem [{\citenamefont {Liu}\ \emph {et~al.}(2010)\citenamefont {Liu},
  \citenamefont {Bedau}, \citenamefont {Backes}, \citenamefont {Katine},
  \citenamefont {Langer},\ and\ \citenamefont {Kent}}]{Liu2010}%
  \BibitemOpen
  \bibfield  {author} {\bibinfo {author} {\bibfnamefont {H.}~\bibnamefont
  {Liu}}, \bibinfo {author} {\bibfnamefont {D.}~\bibnamefont {Bedau}}, \bibinfo
  {author} {\bibfnamefont {D.}~\bibnamefont {Backes}}, \bibinfo {author}
  {\bibfnamefont {J.~A.}\ \bibnamefont {Katine}}, \bibinfo {author}
  {\bibfnamefont {J.}~\bibnamefont {Langer}}, \ and\ \bibinfo {author}
  {\bibfnamefont {A.~D.}\ \bibnamefont {Kent}},\ }\bibfield  {title} {\enquote
  {\bibinfo {title} {Ultrafast switching in magnetic tunnel junction based
  orthogonal spin transfer devices},}\ }\href@noop {} {\bibfield  {journal}
  {\bibinfo  {journal} {Applied Physics Letters}\ }\textbf {\bibinfo {volume}
  {97}},\ \bibinfo {pages} {242510} (\bibinfo {year} {2010})}\BibitemShut
  {NoStop}%
\bibitem [{\citenamefont {Nikonov}\ \emph {et~al.}(2010)\citenamefont
  {Nikonov}, \citenamefont {Bourianoff}, \citenamefont {Rowlands},\ and\
  \citenamefont {Krivorotov}}]{Nikonov2010}%
  \BibitemOpen
  \bibfield  {author} {\bibinfo {author} {\bibfnamefont {Dmitri~E.}\
  \bibnamefont {Nikonov}}, \bibinfo {author} {\bibfnamefont {George~I.}\
  \bibnamefont {Bourianoff}}, \bibinfo {author} {\bibfnamefont {Graham}\
  \bibnamefont {Rowlands}}, \ and\ \bibinfo {author} {\bibfnamefont {Ilya~N.}\
  \bibnamefont {Krivorotov}},\ }\bibfield  {title} {\enquote {\bibinfo {title}
  {Strategies and tolerances of spin transfer torque switching},}\ }\href@noop
  {} {\bibfield  {journal} {\bibinfo  {journal} {Journal of Applied Physics}\
  }\textbf {\bibinfo {volume} {107}},\ \bibinfo {eid} {113910} (\bibinfo {year}
  {2010})}\BibitemShut {NoStop}%
\bibitem [{\citenamefont {Lee}\ \emph {et~al.}(2011)\citenamefont {Lee},
  \citenamefont {Ralph},\ and\ \citenamefont {Buhrman}}]{Lee2011}%
  \BibitemOpen
  \bibfield  {author} {\bibinfo {author} {\bibfnamefont {O.~J.}\ \bibnamefont
  {Lee}}, \bibinfo {author} {\bibfnamefont {D.~C.}\ \bibnamefont {Ralph}}, \
  and\ \bibinfo {author} {\bibfnamefont {R.~A.}\ \bibnamefont {Buhrman}},\
  }\bibfield  {title} {\enquote {\bibinfo {title} {Spin-torque-driven ballistic
  precessional switching with 50 ps impulses},}\ }\href@noop {} {\bibfield
  {journal} {\bibinfo  {journal} {Applied Physics Letters}\ }\textbf {\bibinfo
  {volume} {99}},\ \bibinfo {eid} {102507} (\bibinfo {year}
  {2011})}\BibitemShut {NoStop}%
\bibitem [{\citenamefont {Rowlands}\ \emph {et~al.}(2011)\citenamefont
  {Rowlands}, \citenamefont {Rahman}, \citenamefont {Katine}, \citenamefont
  {Langer}, \citenamefont {Lyle}, \citenamefont {Zhao}, \citenamefont {Alzate},
  \citenamefont {Kovalev}, \citenamefont {Tserkovnyak}, \citenamefont {Zeng},
  \citenamefont {Jiang}, \citenamefont {Galatsis}, \citenamefont {Huai},
  \citenamefont {Amiri}, \citenamefont {Wang}, \citenamefont {Krivorotov},\
  and\ \citenamefont {Wang}}]{Rowlands2011}%
  \BibitemOpen
  \bibfield  {author} {\bibinfo {author} {\bibfnamefont {G.~E.}\ \bibnamefont
  {Rowlands}}, \bibinfo {author} {\bibfnamefont {T.}~\bibnamefont {Rahman}},
  \bibinfo {author} {\bibfnamefont {J.~A.}\ \bibnamefont {Katine}}, \bibinfo
  {author} {\bibfnamefont {J.}~\bibnamefont {Langer}}, \bibinfo {author}
  {\bibfnamefont {A.}~\bibnamefont {Lyle}}, \bibinfo {author} {\bibfnamefont
  {H.}~\bibnamefont {Zhao}}, \bibinfo {author} {\bibfnamefont {J.~G.}\
  \bibnamefont {Alzate}}, \bibinfo {author} {\bibfnamefont {A.~A.}\
  \bibnamefont {Kovalev}}, \bibinfo {author} {\bibfnamefont {Y.}~\bibnamefont
  {Tserkovnyak}}, \bibinfo {author} {\bibfnamefont {Z.~M.}\ \bibnamefont
  {Zeng}}, \bibinfo {author} {\bibfnamefont {H.~W.}\ \bibnamefont {Jiang}},
  \bibinfo {author} {\bibfnamefont {K.}~\bibnamefont {Galatsis}}, \bibinfo
  {author} {\bibfnamefont {Y.~M.}\ \bibnamefont {Huai}}, \bibinfo {author}
  {\bibfnamefont {P.~Khalili}\ \bibnamefont {Amiri}}, \bibinfo {author}
  {\bibfnamefont {K.~L.}\ \bibnamefont {Wang}}, \bibinfo {author}
  {\bibfnamefont {I.~N.}\ \bibnamefont {Krivorotov}}, \ and\ \bibinfo {author}
  {\bibfnamefont {J.-P.}\ \bibnamefont {Wang}},\ }\bibfield  {title} {\enquote
  {\bibinfo {title} {Deep subnanosecond spin torque switching in magnetic
  tunnel junctions with combined in-plane and perpendicular polarizers},}\
  }\href@noop {} {\bibfield  {journal} {\bibinfo  {journal} {Applied Physics
  Letters}\ }\textbf {\bibinfo {volume} {98}},\ \bibinfo {eid} {102509}
  (\bibinfo {year} {2011})}\BibitemShut {NoStop}%
\bibitem [{\citenamefont {Liu}\ \emph {et~al.}(2012)\citenamefont {Liu},
  \citenamefont {Bedau}, \citenamefont {Backes}, \citenamefont {Katine},\ and\
  \citenamefont {Kent}}]{Liu2012}%
  \BibitemOpen
  \bibfield  {author} {\bibinfo {author} {\bibfnamefont {H.}~\bibnamefont
  {Liu}}, \bibinfo {author} {\bibfnamefont {D.}~\bibnamefont {Bedau}}, \bibinfo
  {author} {\bibfnamefont {D.}~\bibnamefont {Backes}}, \bibinfo {author}
  {\bibfnamefont {J.~A.}\ \bibnamefont {Katine}}, \ and\ \bibinfo {author}
  {\bibfnamefont {A.~D.}\ \bibnamefont {Kent}},\ }\bibfield  {title} {\enquote
  {\bibinfo {title} {Precessional reversal in orthogonal spin transfer magnetic
  random access memory devices},}\ }\href@noop {} {\bibfield  {journal}
  {\bibinfo  {journal} {Applied Physics Letters}\ }\textbf {\bibinfo {volume}
  {101}},\ \bibinfo {eid} {032403} (\bibinfo {year} {2012})}\BibitemShut
  {NoStop}%
\bibitem [{\citenamefont {Park}\ \emph {et~al.}(2013)\citenamefont {Park},
  \citenamefont {Ralph},\ and\ \citenamefont {Buhrman}}]{Park2013}%
  \BibitemOpen
  \bibfield  {author} {\bibinfo {author} {\bibfnamefont {Junbo}\ \bibnamefont
  {Park}}, \bibinfo {author} {\bibfnamefont {D.~C.}\ \bibnamefont {Ralph}}, \
  and\ \bibinfo {author} {\bibfnamefont {R.~A.}\ \bibnamefont {Buhrman}},\
  }\bibfield  {title} {\enquote {\bibinfo {title} {Fast deterministic switching
  in orthogonal spin torque devices via the control of the relative spin
  polarizations},}\ }\href@noop {} {\bibfield  {journal} {\bibinfo  {journal}
  {Applied Physics Letters}\ }\textbf {\bibinfo {volume} {103}},\ \bibinfo
  {eid} {252406} (\bibinfo {year} {2013})}\BibitemShut {NoStop}%
\bibitem [{\citenamefont {Ye}\ \emph {et~al.}(2014)\citenamefont {Ye},
  \citenamefont {Gopman}, \citenamefont {Rehm}, \citenamefont {Backes},
  \citenamefont {Wolf}, \citenamefont {Ohki}, \citenamefont {Kirichenko},
  \citenamefont {Vernik}, \citenamefont {Mukhanov},\ and\ \citenamefont
  {Kent}}]{Ye2014}%
  \BibitemOpen
  \bibfield  {author} {\bibinfo {author} {\bibfnamefont {L.}~\bibnamefont
  {Ye}}, \bibinfo {author} {\bibfnamefont {D.~B.}\ \bibnamefont {Gopman}},
  \bibinfo {author} {\bibfnamefont {L.}~\bibnamefont {Rehm}}, \bibinfo {author}
  {\bibfnamefont {D.}~\bibnamefont {Backes}}, \bibinfo {author} {\bibfnamefont
  {G.}~\bibnamefont {Wolf}}, \bibinfo {author} {\bibfnamefont {T.}~\bibnamefont
  {Ohki}}, \bibinfo {author} {\bibfnamefont {A.~F.}\ \bibnamefont
  {Kirichenko}}, \bibinfo {author} {\bibfnamefont {I.~V.}\ \bibnamefont
  {Vernik}}, \bibinfo {author} {\bibfnamefont {O.~A.}\ \bibnamefont
  {Mukhanov}}, \ and\ \bibinfo {author} {\bibfnamefont {A.~D.}\ \bibnamefont
  {Kent}},\ }\bibfield  {title} {\enquote {\bibinfo {title} {Spin-transfer
  switching of orthogonal spin-valve devices at cryogenic temperatures},}\
  }\href@noop {} {\bibfield  {journal} {\bibinfo  {journal} {Journal of Applied
  Physics}\ }\textbf {\bibinfo {volume} {115}},\ \bibinfo {eid} {17C725}
  (\bibinfo {year} {2014})}\BibitemShut {NoStop}%
\bibitem [{\citenamefont {Houssameddine}\ \emph {et~al.}(2007)\citenamefont
  {Houssameddine}, \citenamefont {Ebels}, \citenamefont {Dela{\"e}t},
  \citenamefont {Rodmacq}, \citenamefont {Firastrau}, \citenamefont
  {Ponthenier}, \citenamefont {Brunet}, \citenamefont {Thirion}, \citenamefont
  {Michel}, \citenamefont {Prejbeanu-Buda}, \citenamefont {Cyrille},
  \citenamefont {Redon},\ and\ \citenamefont {Dieny}}]{Houssameddine2007}%
  \BibitemOpen
  \bibfield  {author} {\bibinfo {author} {\bibfnamefont {D.}~\bibnamefont
  {Houssameddine}}, \bibinfo {author} {\bibfnamefont {U.}~\bibnamefont
  {Ebels}}, \bibinfo {author} {\bibfnamefont {B.}~\bibnamefont {Dela{\"e}t}},
  \bibinfo {author} {\bibfnamefont {B.}~\bibnamefont {Rodmacq}}, \bibinfo
  {author} {\bibfnamefont {I.}~\bibnamefont {Firastrau}}, \bibinfo {author}
  {\bibfnamefont {F.}~\bibnamefont {Ponthenier}}, \bibinfo {author}
  {\bibfnamefont {M.}~\bibnamefont {Brunet}}, \bibinfo {author} {\bibfnamefont
  {C.}~\bibnamefont {Thirion}}, \bibinfo {author} {\bibfnamefont {J-P.}\
  \bibnamefont {Michel}}, \bibinfo {author} {\bibfnamefont {L.}~\bibnamefont
  {Prejbeanu-Buda}}, \bibinfo {author} {\bibfnamefont {M.-C.}\ \bibnamefont
  {Cyrille}}, \bibinfo {author} {\bibfnamefont {O.}~\bibnamefont {Redon}}, \
  and\ \bibinfo {author} {\bibfnamefont {B.}~\bibnamefont {Dieny}},\ }\bibfield
   {title} {\enquote {\bibinfo {title} {Spin-torque oscillator using a
  perpendicular polarizer and a planar free layer},}\ }\href@noop {} {\bibfield
   {journal} {\bibinfo  {journal} {Nature Materials}\ }\textbf {\bibinfo
  {volume} {6}},\ \bibinfo {pages} {447--453} (\bibinfo {year}
  {2007})}\BibitemShut {NoStop}%
\bibitem [{\citenamefont {Ebels}\ \emph {et~al.}(2008)\citenamefont {Ebels},
  \citenamefont {Houssameddine}, \citenamefont {Firastrau}, \citenamefont
  {Gusakova}, \citenamefont {Thirion}, \citenamefont {Dieny},\ and\
  \citenamefont {Buda-Prejbeanu}}]{Ebels2008}%
  \BibitemOpen
  \bibfield  {author} {\bibinfo {author} {\bibfnamefont {U.}~\bibnamefont
  {Ebels}}, \bibinfo {author} {\bibfnamefont {D.}~\bibnamefont
  {Houssameddine}}, \bibinfo {author} {\bibfnamefont {I.}~\bibnamefont
  {Firastrau}}, \bibinfo {author} {\bibfnamefont {D.}~\bibnamefont {Gusakova}},
  \bibinfo {author} {\bibfnamefont {C.}~\bibnamefont {Thirion}}, \bibinfo
  {author} {\bibfnamefont {B.}~\bibnamefont {Dieny}}, \ and\ \bibinfo {author}
  {\bibfnamefont {L.~D.}\ \bibnamefont {Buda-Prejbeanu}},\ }\bibfield  {title}
  {\enquote {\bibinfo {title} {Macrospin description of the perpendicular
  polarizer-planar free-layer spin-torque oscillator},}\ }\href@noop {}
  {\bibfield  {journal} {\bibinfo  {journal} {Phys. Rev. B}\ }\textbf {\bibinfo
  {volume} {78}},\ \bibinfo {pages} {024436} (\bibinfo {year}
  {2008})}\BibitemShut {NoStop}%
\bibitem [{\citenamefont {Firastrau}\ \emph {et~al.}(2008)\citenamefont
  {Firastrau}, \citenamefont {Gusakova}, \citenamefont {Houssameddine},
  \citenamefont {Ebels}, \citenamefont {Cyrille}, \citenamefont {Delaet},
  \citenamefont {Dieny}, \citenamefont {Redon}, \citenamefont {Toussaint},\
  and\ \citenamefont {Buda-Prejbeanu}}]{Firastrau2008}%
  \BibitemOpen
  \bibfield  {author} {\bibinfo {author} {\bibfnamefont {I.}~\bibnamefont
  {Firastrau}}, \bibinfo {author} {\bibfnamefont {D.}~\bibnamefont {Gusakova}},
  \bibinfo {author} {\bibfnamefont {D.}~\bibnamefont {Houssameddine}}, \bibinfo
  {author} {\bibfnamefont {U.}~\bibnamefont {Ebels}}, \bibinfo {author}
  {\bibfnamefont {M.-C.}\ \bibnamefont {Cyrille}}, \bibinfo {author}
  {\bibfnamefont {B.}~\bibnamefont {Delaet}}, \bibinfo {author} {\bibfnamefont
  {B.}~\bibnamefont {Dieny}}, \bibinfo {author} {\bibfnamefont
  {O.}~\bibnamefont {Redon}}, \bibinfo {author} {\bibfnamefont {J.-Ch.}\
  \bibnamefont {Toussaint}}, \ and\ \bibinfo {author} {\bibfnamefont {L.~D.}\
  \bibnamefont {Buda-Prejbeanu}},\ }\bibfield  {title} {\enquote {\bibinfo
  {title} {Modeling of the perpendicular polarizer-planar free layer spin
  torque oscillator: Micromagnetic simulations},}\ }\href@noop {} {\bibfield
  {journal} {\bibinfo  {journal} {Phys. Rev. B}\ }\textbf {\bibinfo {volume}
  {78}},\ \bibinfo {pages} {024437} (\bibinfo {year} {2008})}\BibitemShut
  {NoStop}%
\bibitem [{\citenamefont {Pinna}\ \emph {et~al.}(2013)\citenamefont {Pinna},
  \citenamefont {Kent},\ and\ \citenamefont {Stein}}]{Pinna2013}%
  \BibitemOpen
  \bibfield  {author} {\bibinfo {author} {\bibfnamefont {D.}~\bibnamefont
  {Pinna}}, \bibinfo {author} {\bibfnamefont {A.~D.}\ \bibnamefont {Kent}}, \
  and\ \bibinfo {author} {\bibfnamefont {D.~L.}\ \bibnamefont {Stein}},\
  }\bibfield  {title} {\enquote {\bibinfo {title} {Thermally assisted
  spin-transfer torque dynamics in energy space},}\ }\href {\doibase
  10.1103/PhysRevB.88.104405} {\bibfield  {journal} {\bibinfo  {journal} {Phys.
  Rev. B}\ }\textbf {\bibinfo {volume} {88}},\ \bibinfo {pages} {104405}
  (\bibinfo {year} {2013})}\BibitemShut {NoStop}%
\bibitem [{\citenamefont {Slonczewski}(1996)}]{Slonczewski1996}%
  \BibitemOpen
  \bibfield  {author} {\bibinfo {author} {\bibfnamefont {J.~C.}\ \bibnamefont
  {Slonczewski}},\ }\bibfield  {title} {\enquote {\bibinfo {title}
  {Current-driven excitation of magnetic multilayers},}\ }\href@noop {}
  {\bibfield  {journal} {\bibinfo  {journal} {Journal of Magnetism and Magnetic
  Materials}\ }\textbf {\bibinfo {volume} {159}},\ \bibinfo {pages} {L1--L7}
  (\bibinfo {year} {1996})}\BibitemShut {NoStop}%
\bibitem [{\citenamefont {{Pinna}}\ \emph {et~al.}(2014)\citenamefont
  {{Pinna}}, \citenamefont {{Stein}},\ and\ \citenamefont
  {{Kent}}}]{Pinna2014}%
  \BibitemOpen
  \bibfield  {author} {\bibinfo {author} {\bibfnamefont {D.}~\bibnamefont
  {{Pinna}}}, \bibinfo {author} {\bibfnamefont {D.~L.}\ \bibnamefont
  {{Stein}}}, \ and\ \bibinfo {author} {\bibfnamefont {A.~D.}\ \bibnamefont
  {{Kent}}},\ }\bibfield  {title} {\enquote {\bibinfo {title} {{Spin Torque
  Oscillators with Thermal Noise: A Constant Energy Orbit Approach}},}\
  }\href@noop {} {\bibfield  {journal} {\bibinfo  {journal} {ArXiv e-prints}\ }
  (\bibinfo {year} {2014})},\ \Eprint {http://arxiv.org/abs/1405.0731}
  {arXiv:1405.0731:1405.0731 [cond-mat.mes-hall]} \BibitemShut {NoStop}%
\end{thebibliography}
\end{document}